\documentclass[aps,prl,reprint,showpacs,preprintnumbers,amsmath,amssymb,floatfix,a4paper,superscriptaddress]{revtex4-1}

\usepackage[T1]{fontenc}
\usepackage[utf8x]{inputenc}

\usepackage{lmodern}
\usepackage{graphicx}
\usepackage{dcolumn}
\usepackage{bm}
\usepackage{color}
\usepackage{isotope,xspace}
\usepackage[squaren,thinspace,mediumqspace,Gray]{SIunits}

\usepackage{xcolor}
\definecolor{medium-blue}{rgb}{0,0,0.5}
\usepackage[colorlinks=true,final,unicode]{hyperref}
\hypersetup{
	pdftitle={Measurement of the weak axial-vector coupling constant in the decay of free neutrons using a pulsed cold neutron beam},
	pdfauthor={B. Märkisch, H. Mest, H. Saul, X. Wang,  H. Abele, D. Dubbers, M. Klopf, A. Petoukhov, C. Roick, T. Soldner, D. Werder},
    colorlinks, linkcolor={medium-blue},
    citecolor={medium-blue}, urlcolor={medium-blue}
}

\newcolumntype{d}{D{.}{.}{-1}}

\newcommand{\perkeo}{\textsc{Perkeo\ III}\xspace}

\newcommand{\sn}{\isotope[113]{Sn}\xspace}
\newcommand{\bi}{\isotope[207]{Bi}\xspace}
\newcommand{\ce}{\isotope[139]{Ce}\xspace}

\newcommand{\cs}{\isotope[137]{Cs}\xspace}

\newcommand{\lif}{\isotope[6]{LiF}\xspace}
\newcommand{\he}{\isotope[3]{He}\xspace}

\newcommand{\heidelberg}{Physikalisches Institut, Universität Heidelberg, Im~Neuenheimer~Feld~226, 69120 Heidelberg, Germany}
\newcommand{\ill}{Institut Laue-Langevin, 71 avenue des Martyrs, CS 20156, 38042 Grenoble Cedex 9, France}
\newcommand{\ati}{Technische Universität Wien, Atominstitut, Stadionallee~2, 1020 Wien, Austria}
\newcommand{\tum}{Physik-Department, Technische Universität München, James-Franck-Straße 1, 85748 Garching, Germany}
\newcommand{\frm}{Forschungs-Neutronenquelle Heinz Maier-Leibnitz (FRM~II), Technische Universität München, Lichtenbergstraße 1, 85748 Garching, Germany}

\begin{document}

%\preprint{2018019}

\title{Measurement of the Weak Axial-Vector Coupling Constant in the Decay of Free Neutrons Using a Pulsed Cold Neutron Beam}

\author{B.~Märkisch}
\email{maerkisch@ph.tum.de}
\affiliation{\tum}
\affiliation{\heidelberg}

\author{H.~Mest}
\affiliation{\heidelberg}

\author{H.~Saul}
\affiliation{\tum}
\affiliation{\ati}
\affiliation{\frm}

\author{X.~Wang}
\affiliation{\tum}
\affiliation{\ati}

\author{H.~Abele}
\email{abele@ati.ac.at}
\affiliation{\tum}
\affiliation{\heidelberg}
\affiliation{\ati}

\author{D.~Dubbers}
\affiliation{\heidelberg}

\author{M.~Klopf}
\affiliation{\ati}

\author{A.~Petoukhov}
\affiliation{\ill}

\author{C.~Roick}
\affiliation{\tum}
\affiliation{\heidelberg}

\author{T.~Soldner}
\affiliation{\ill}

\author{D.~Werder}
\affiliation{\heidelberg}

\begin{abstract}
We present a precision measurement of the axial-vector coupling constant $g_A$ in the decay of polarized free neutrons.  For the first time, a pulsed cold neutron beam was used for this purpose. By this method, leading sources of systematic uncertainty are suppressed.  From the electron spectra we obtain $\lambda = g_A/g_V = -1.27641(45)_\mathrm{stat}(33)_\mathrm{sys}$ which confirms recent measurements with improved precision. This corresponds to a value of the parity violating beta asymmetry parameter of $A_0 = -0.11985(17)_\mathrm{stat}(12)_\mathrm{sys}$. We discuss implications on the CKM matrix element $V_{ud}$ and derive a limit on left-handed tensor interaction.
\end{abstract}

\date{\today}


\maketitle

%\paragraph{Introduction}
Measurements in muon and neutron decay link weak leptonic and semileptonic decays to the underlying 
Lagrangian of the standard model with coupling constants based on the $SU(3)_c \times SU(2)_L \times U(1)$ gauge structure \cite{Glashow61,Weinberg67,Salam68}. The Fermi coupling constant $G_F$ measured in muon decay \cite{Webber11} provides a low energy value for the weak coupling and was historically used to predict the masses of the $W$ and $Z$ bosons \cite{Sirlin80,Marciano80,Llewellyn81}.  The Lagrange density in semileptonic neutron decay includes a hadronic vector $(V)$ and an axial vector current $(A)$.
The vector coupling constant for quarks $g_V$ is related to that for leptons via $g_V = G_F\,V_{ud}$, with the matrix element $V_{ud}$ of the Cabbibo-Kobayashi-Maskawa (CKM) quark mixing matrix.  The axial vector current is renormalized by the strong interaction at low energy.  This is quantified by the parameter $\lambda = g_A/g_V$, the ratio of the axial vector and vector coupling constants. If the weak interaction is invariant under time reversal, the parameter $\lambda$ is real. 

The most precise experimental determination of the parameter $\lambda$ is from the beta asymmetry in neutron decay but older experimental results \cite{Bopp86,Yero97,Liaud97} are not consistent with newer experiments \cite{Mund13,Brown18} with smaller systematic corrections. This discrepancy prevails since the first result of \textsc{Perkeo~II} \citep{Abele97} and one aim of this work is to clarify the situation.

The ratio of coupling constants $\lambda$ is important in many fields \cite{Dubbers11}:  it enters in the prediction for the energy consumption in the sun via the primary reaction in the pp-chain and the solar neutrino flux, the production of light elements in primordial nucleosynthesis, and neutron star formation. $\lambda$ is also used for the calibration of neutrino detectors and as an input for the unitarity check of the quark-mixing CKM matrix.  Within the standard model the CKM matrix is unitary, and unitarity tests are sensitive tools for searches for new physics \cite{Cirigliano10}. Experimentally, unitarity is verified at the $10^{-4}$ level for the first row of the matrix using nuclear decays~\cite{Hardy15}. The possible observation of non-standard model couplings is another example how new physics could emerge and is typically tested in the framework of effective field theories, for recent reviews and surveys see \cite{Gonzalez-Alonso18,Holstein14,Bhattacharya12,Cirigliano13,Vos15,Dubbers11}.

In neutron decay, the probability that an electron is emitted with angle $\vartheta$ with respect to the neutron spin polarization vector $\bm{P} = \langle\bm{\sigma}\rangle/\sigma$ is \cite{Jackson57}
\begin{equation}
W(\vartheta) = 1 + \frac{v}{c}PA\cos\vartheta,\label{eq:angulardist}
\end{equation}
where $v$ is the electron velocity. ${A}$ is the parity violating beta asymmetry parameter which depends on $\lambda$.

%\paragraph{In this letter:}
In this letter we present the first determination of $\lambda$ from a measurement of the beta asymmetry using a pulsed neutron beam. The method is described in Ref.\ \cite{Maerkisch09,Maerkisch14} and effectively eliminates or controls leading sources of systematic uncertainty: beam related background, edge effects \cite{Dubbers14}, and the magnetic mirror effect. Yet, one order of magnitude more data was collected compared to the previously most precise experiment \cite{Mund13}. In order to ensure a \emph{blind} analysis, the beta decay data and two major systematic corrections (neutron beam polarization and magnetic mirror effect) were analyzed separately by independent teams. Results were combined once the analyses were complete.

\begin{figure*}[tb]
\includegraphics[width=\textwidth]{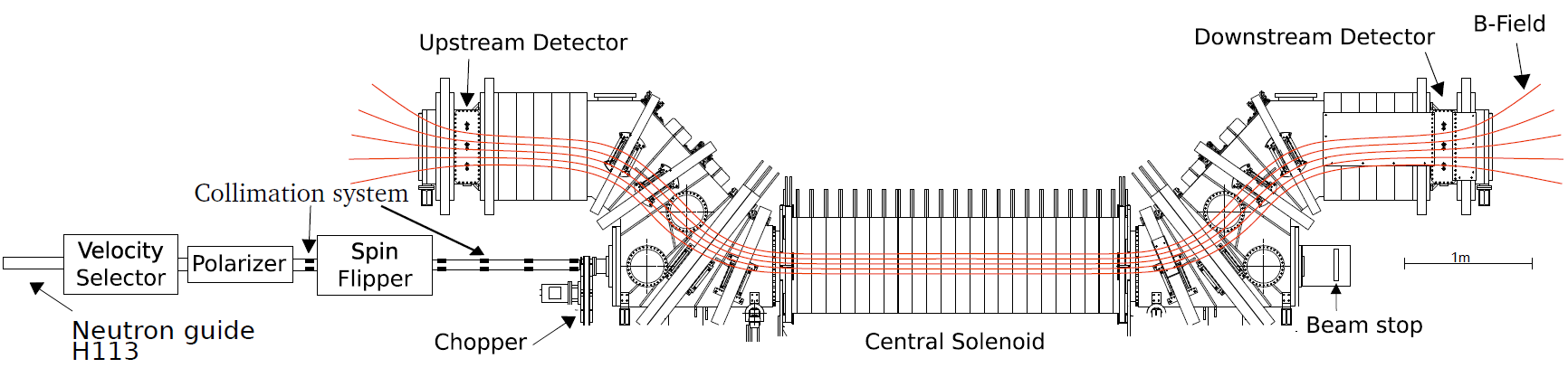}
\caption{Schematic of the spectrometer setup \perkeo installed at PF1B. The magnetic field is indicated in red. The cold neutron beam from the reactor source enters the beam line on the left hand side. Picture from \cite{Mest11}.}
 \label{fig:overview}
\end{figure*}

%\paragraph{Spectrometer:}
The spectrometer \perkeo \cite{Maerkisch09,Maerkisch14} is the successor to the spectrometer \textsc{Perkeo~II}, which was used to measure the beta asymmetry parameter $A$ \cite{Abele97,Abele02,Mund13}, the neutrino asymmetry parameter $B$ \cite{Kreuz05a,Schumann07}, and the proton asymmetry parameter $C$ \cite{Schumann08a} in polarized neutron beta decay. The main component of the spectrometer is an $\unit{8}{m}$ long magnet system, consisting of 54 short conventional coils with rectangular cross section. The shape of the magnetic field and the beam line set-up are shown in Fig.~\ref{fig:overview}. Electrons from neutron decay in the (nearly) homogeneous field in the center of the spectrometer are separated from the neutron beam and guided to two detectors installed upstream and downstream. By using a pulsed beam neutrons are temporarily stored in-flight, fully contained in the decay volume without any contact to material. The solid angle coverage of the detectors is thus truly $2\!\times\! 2\pi$, without edge effects. The magnetic field has its maximum of $B_{\textrm{max}} = \unit{152.5}{mT}$ in the center of the active region. The field decreases towards the detectors to reduce backscatter effects from the detectors. Details on electron backscatter suppression can be found in Refs.~\cite{Abele97,Abele02,Roick18c}.

\begin{figure}[b]
\centering
\includegraphics{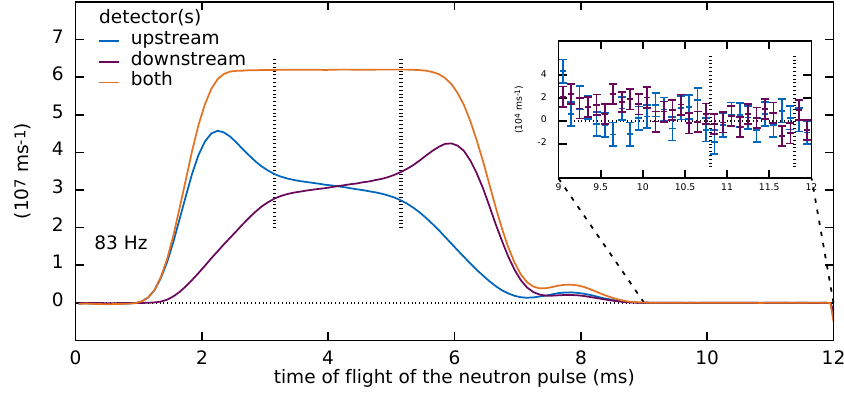}
\caption{Number of decay events in the electron energy window $300\ldots\unit{700}{keV}$ as a function of time of flight of the neutron pulse through the spectrometer for a chopper frequency of $\unit{83}{Hz}$. Background is subtracted. The insert zooms in on the background time window. The shape of the curves for the detectors is caused by the magnetic mirror effect. The vertical lines indicate the time windows used for signal and background extraction.}
 \label{fig:ToF}
\end{figure}

%\paragraph{Beamline:}
The spectrometer was installed at the PF1B cold neutron beam position at the Institut Laue-Langevin (ILL) \cite{Haese02,Abele06}. Neutrons moderated by a cold source are transported to the beam site via a supermirror guide. The capture flux density measured at the guide's exit was $\Phi=\unit{2\times 10^{10}}{s^{-1}cm^{-2}}$. A Dornier neutron velocity selector \cite{Wagner92} allowed only neutrons within a wavelength band of $\unit{4.5\ldots 5.5}{\angstrom}$ to enter the instrument's beam line. The neutron beam was subsequently polarized by a single supermirror (SM) coated bender polarizer \cite{Soldner02}. The adjustment of the polarizer was optimized for transmission which also yields a rather symmetrical beam intensity and polarization. An adiabatic fast passage flipper allowed to reverse the neutron spin direction. A series of five apertures made out of \lif were used to shape the neutron beam. Before the neutron beam entered the \perkeo spectrometer, a rotating disc chopper with a maximum frequency of $\unit{6000}{rpm}$ was used to pulse the beam as described in Ref.~\cite{Maerkisch09}. \lif ceramics embedded in the chopper disc made of fiber reinforced plastic were used to absorb the neutron beam. The geometrical opening of the chopper disc was $\unit{7.3}{\%}$. After passing the spectrometer, remaining neutrons were dumped in a \isotope[10]{B}$_{4}$C beam stop. Data from neutron decay were taken at two different frequencies of the chopper, $\unit{83}{Hz}$ and $\unit{94}{Hz}$. Both datasets have similar size. The number of detected decay events as a function of time after the chopper opens is shown in Fig.~\ref{fig:ToF}.

When used with a polarized continuous neutron beam, \perkeo can detect up to $\unit{5 \times 10^4}{decay\ events/s}$ \cite{Maerkisch09}. In the current set-up -- using a pulsed neutron beam -- the instantaneous decay rate was reduced to $\unit{2 \times 10^3}{s^{-1}}$ during the signal time interval and $\unit{1.4\times 10^2}{s^{-1}}$ on time average. In total $6\times 10^8$ neutron decay events were used for the extraction of the results.

%\paragraph{Polarization:}
The polarization of the neutron beam was measured for several rotational speeds of the velocity selector using opaque \he spin filter cells. Regular in-situ flipping of the \he spins served to separate neutron flipping efficiency and polarization. Flat and parallel neutron entrance and exit windows of the filter cells and homogeneous neutron detectors assured a correct spatial averaging over the rectangular aperture used to define the sensitive area. The beam averaged polarization was determined from polarization and intensity scans with step sizes equal to the dimensions of this aperture, avoiding interpolation uncertainties. Scans performed from a shifted reference point yielded consistent results. The derived systematic uncertainty of spatial averaging of $3.4\times 10^{-4}$ is dominated by the limited statistical precision of these tests. Measurements in front and behind the \perkeo spectrometer gave consistent results. Later measurements behind the spectrometer, in between and after the beta decay runs, were used to derive the limit of $3.9\times 10^{-4}$ on a potential time variation of the beam polarization. The third leading systematic uncertainty stems from the measured correction for neutron depolarization by the spectrometer exit window of $2.7(2.2)\times 10^{-4}$. The background was determined from regions outside the neutron pulse in the time of flight spectra and results in a correction of $2.3(9)\times 10^{-4}$. All other individual corrections and uncertainties, including that for the averaging over the wavelength band, are below $1\times 10^{-4}$. The resulting average neutron polarization of the pulses at the reference selector speed is $P=0.99100(60)$ and the flipping efficiency is $> 0.99964\ (68.3\%\ \mathrm{C.L.})$. The combined error of these two effects on the asymmetry is $6.4\times 10^{-4}$. Since we do not use two SM bender polarizers in crossed (X-SM) geometry \cite{Kreuz05b} the resulting neutron polarization is lower than in Ref.~\cite{Mund13}, but the accuracy is improved by about a factor of two due to the limited wavelength band and improved systematic uncertainties \cite{Soldner11,Klauser12,Klauser13,Klauser16}.

%\paragraph{Electron detector:}
Electrons were detected using two plastic scintillators of size $43\times \unit{45}{cm^2}$ which were each read out by six fine mesh photomultiplier tubes. Mono-energetic conversion electron sources (\ce, \sn, \cs, and \bi) were used to calibrate and characterize the detectors. The sources were supported by thin carbon foil backings ($<\unit{20}{µg/cm^2}$) and could be moved in two dimensions across the cross-section of the neutron beam in the decay volume. Detector drifts were checked and accounted for with a single source every hour. Twice a day, a full calibration with all sources was performed. The non-linearity of the detector systems was determined using the calibration sources and off-line measurements of the electronics and the scintillator. While calibration with sources works very well at energies larger than $\unit{200}{keV}$, future measurements will also employ a time-of-flight technique to improve calibration at low energies \cite{Dubbers08,Roick18}.
%\paragraph{Uniformity:}
The uniformity of the detectors was measured about once a week. As a major improvement over previous \textsc{Perkeo} measurements, the variation of the detector amplitude over the area covered by decay electrons was $<\pm\unit{2.5}{\%}$. The effect of the detector non-uniformity on our result is dominated by the dependence of the magnetic point-spread function on the beta asymmetry \cite{Dubbers14, Dubbers15} and the difference in detector coverage of beta asymmetry and calibration measurements. After unblinding we corrected the geometrical coverage of the detector using a previously unapplied analysis, which shifted the result by less than $\Delta \lambda / 10$. The corresponding corrections are calculatued using photon transport simulations performed with GEANT4 \cite{Geant4}. The resulting detector model fits the measured uniformity under consideration of the electron point-spread. The overall non-uniformity correction also includes the effect of energy-loss in the carbon foil backings which only needs to be considered for calibration measurements.

%\paragraph{Unrecognised Backscattering}
Despite the suppression of backscattering from the detectors by the magnetic field, about \unit{6}{\%} of the electrons are reflected onto the opposite detector. Full energy reconstruction is possible as both detectors are always read out simultaneously. If the deposited energy during the first incidence is not sufficient for a trigger, the wrong emission direction would be assigned to the event. The correction for this effect is calculated based on Monte Carlo backscattering simulations using GEANT4, which have been verified using measured backscattering data \cite{Roick18c}.

%\paragraph{Background:}
Neutron decay data was only taken while the neutron pulse was fully contained in the homogeneous part of the magnetic field and the chopper was closed. Background was measured at the end of every chopper cycle when the neutron beam had been fully absorbed by the neutron beam dump, see Fig.~\ref{fig:ToF}.
Signal and background are thus measured under the same condition, i.e. with the chopper closed. The time dependence of the ambient background and that created by the closed chopper was checked using \he neutron counter tubes, NaI $\gamma$-detectors and the \textsc{Perkeo} detector systems. Effects were found to be smaller than $\Delta A/A = 2\times 10^{-4}$. In previous experiments \cite{Mund13} a significant amount of data had to be discarded due to the variation of background from external sources like neighboring instruments. In this experiment these sources were tracked without significant delay. The uncertainty on the background measurement was reduced by more than an order of magnitude.
The two datasets with different chopper frequencies were used to investigate possible beam dependent background.  The signal-to-background ratio in the fit region was better than $4:1$.

\begin{figure}[htb]
 \centering
 \includegraphics{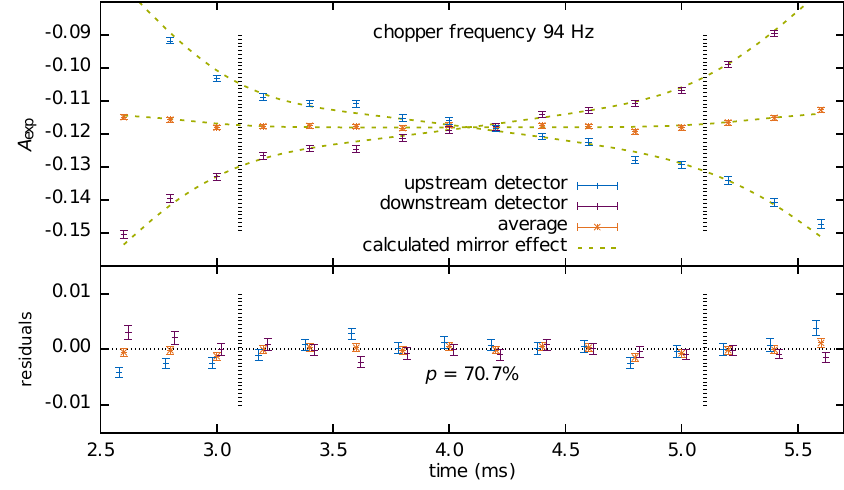}
 \caption{Measured asymmetry $A_\mathrm{exp}$ for both detectors as a function of time of flight of the neutron pulse. When averaging over both detectors, the dominant effect of the magnetic mirror effect cancels and only a small correction remains, see Tab.~\ref{tab:budget}. The vertical dashed lines indicate the time window used for the asymmetry analysis, the dashed lines indicate the calculated mirror effect.}
 \label{fig:Asym-ToF}
\end{figure}

\begin{table}[t]
\centering
\begin{ruledtabular}
\begin{tabular}{l c c}
\textbf{\hphantom{aa}Effect on} & \textbf{Relative}                           & \textbf{Relative}\\
\textbf{Asymmetry \emph{A}}     & \textbf{\hphantom{a}Correction\hphantom{a}} &\textbf{\hphantom{a}Uncertainty\hphantom{a}} \\
                                & $\left(10^{-4}\right)$                      & $\left(10^{-4}\right)$  \\[0.15em]\hline\\[-0.9em]
Neutron beam				&				& \\
\quad Polarization and		& $90.7$		& $6.4$ \\
\quad Spinflip efficiency	&   	 		& \\
Background					&&\\
\quad Time variation		& $-0.8$ 		& $0.8$\\
\quad Chopper				& $-1.9$ 		& $0.7$\\
Electrons					&&\\
\quad Magn.\ mirror effect	& $46.1$ 		& $4.5$\\
\quad Undetected backscattering & $5.0$	 	& $1.5$\\
\quad Lost backscatter energy 	& $0$	 	& $1.4$\\
Electron detector 			&&\\
\quad Deadtime 				& $(5)$\textsuperscript{*}& $0.35$\\
\quad Temporal stability 	&		 		& $3.7$\\
\quad Non-uniformity		& $4.2$ 		& $2.1$\\
\quad Non-linearity			& $-1$\textsuperscript{*}& $4$\\
\quad Calibration (input data)	&    			& $1$\\
Theory			 			&&\\
\quad Ext.~radiative corr.	& $(-10)$\textsuperscript{*}& $1$\\[0.15em]\hline\\[-0.9em]
Total systematics			&  $138.1$ 		& $10.3$\\
Statistical uncertainty		&				& $14.0$ \\[0.15em]\hline\\[-0.9em]
\textbf{Total} 				&				& $\mathbf{17.4}$\\
\end{tabular}
\end{ruledtabular}
\caption{\label{tab:budget}
Summary of corrections to the measured experimental asymmetry and uncertainties. All quantities are given as fractions $\Delta A/A$ of the asymmetry parameter. The fit parameter actually is $\lambda$, but we list corrections on $A$ for comparability with earlier measurements.\\
\textsuperscript{*}Already included in result Eq.~\eqref{eq:final-result}: measured by the data acquisition system or included in the fit function.}
\end{table}

%\paragraph{Mirror effect:}
Electrons are reflected on an increasing magnetic field if the opening angle of gyration with respect to the magnetic field exceeds the critical angle $\theta_c = \arcsin \sqrt{^{B_1}/_{B_0}}$, were $B_0$ is the maximum of the magnetic field and $B_1$ is the field at the place of the neutron decay. This modifies the solid angle coverage of the two detectors and hence changes the measured asymmetry. This expected \emph{mirror effect} is calculated from measurements of the shape of the neutron cloud in space and time and magnetic field distributions. The field in the active region was measured with Hall probes on a $3\times 3\times 27$ grid with a spacing of $\unit{10}{cm}$. Measurements were performed before and after the neutron decay measurements and with different sensors and yield consistent results. The neutron pulse shape in space and time was measured at several positions downstream of the chopper using different detection techniques ($(n,\gamma)$ reactions in aluminum foils or absorbtion in boron and gamma detectors, copper foil activation and scanning with neutron detectors). Results are consistent with each other and agree with calculations and McStas \cite{Lefmann1999,Willendrup2004} simulations based on the geometrical properties of selector and chopper. Most of the mirror effect on the asymmetry $A$ cancels by averaging the results of the upstream and downstream detectors, which can be seen in Fig.~\ref{fig:Asym-ToF}. The remaining average correction of both detectors on the time-averaged asymmetry is $(49.7 \pm 4.5)\times 10^{-4}$ for the $\unit{83}{Hz}$ and $(42.5 \pm 4.5)\times 10^{-4}$ for the $\unit{94}{Hz}$ chopper frequencies. We note that this correction is independent of the electron energy.

%\paragraph{Results:}
The electron spectra for both spin states and detectors ($N_i^{\uparrow\downarrow}(E)$, $i=1,2$) were used to obtain the experimental asymmetry, which is directly related to the asymmetry parameter $A$ of Eq.~\eqref{eq:angulardist}. Omitting some further correction terms, we obtain:
\begin{align}
 A_{\text{exp},i}(E)=\frac{N_i^\uparrow(E)\!-\!N_i^\downarrow(E)}{N_i^\uparrow(E)\!+\!N_i^\downarrow(E)}=\frac{1}{2}\frac{v}{c} A P\frac{M}{1\pm k}\,,
 \label{eq:exp-asym}
\end{align}
where the factor $\langle\cos\theta\rangle=1/2$ results from solid angle integration of Eq.~\eqref{eq:angulardist}. The neutron polarization $P$, the magnetic mirror correction $M/(1\pm k)$ \cite{Raven95} and the beta decay data were all analyzed by independent teams.

The asymmetry parameter $A$ includes order $1\%$ corrections for weak magnetism, $g_V-g_A$ interference, and nucleon recoil \cite{Wilkinson82}. For $\lambda$ real, the uncorrected asymmetry $A_0$ is given by
\begin{equation}
A_0=-2\frac{\lambda(\lambda+1)}{1+3\lambda^2}.
\end{equation}
An additional small radiative correction \cite{Shann71, Glueck92, Ivanov13a} of order $0.1\%$ must be applied.

\begin{figure}[t]
 \centering
 \includegraphics{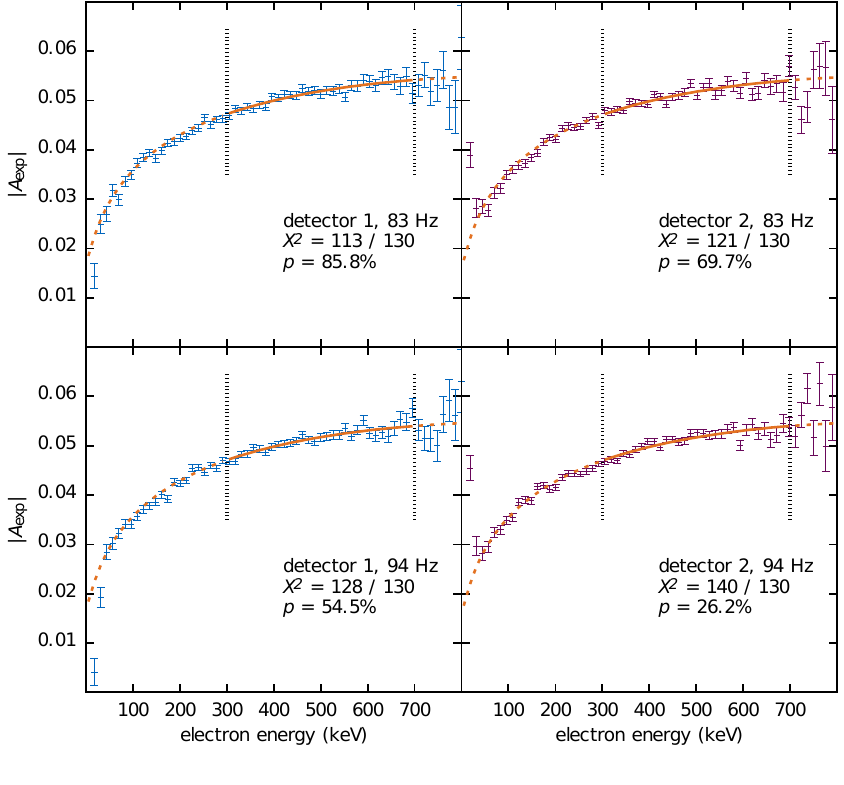}
 \caption{The experimental asymmetry $\left|A_{\mathrm{exp},i}(E_e)\right|$ and fits with a single free parameter $\lambda$ for both detectors and both chopper frequencies. The solid part of the line indicates the fit interval which was chosen to balance statistics and effects due to detector non-linearity and unrecognized backscattering.}
 \label{fig:exp-asym}
\end{figure}

Fig.~\ref{fig:exp-asym} shows fits to the experimental asymmetry of the available four datasets with the ratio of coupling constants $\lambda$ as single free parameter. We note that the calibrations of the detectors were obtained using calibration sources only instead of relying on electron spectra from neutron beta decay. The result is independent of the choice of the fit range within the statistical error. The fit range of $\approx \unit{300}{keV}\ldots\unit{700}{keV}$ was chosen in order to optimize the combined systematic and statistical uncertainties.

We apply remaining corrections listed in Table~\ref{tab:budget} and obtain the final result:
\begin{align}
\lambda &= -1.27641(45)_\mathrm{stat}(33)_\mathrm{sys}  \nonumber \\
		&= -1.27641(56) \label{eq:final-result} \\
A 		&= -0.11985(17)_\mathrm{stat}(12)_\mathrm{sys} \nonumber \\
		&= -0.11985(21). \nonumber
\end{align}
The result of this blinded analysis confirms recent measurements using cold and ultra-cold neutrons \cite{Mund13,Brown18} with $2.5$ times higher precision compared to \cite{Mund13}. We note that all recent measurements have much smaller corrections on the percent level only compared to older measurements, see also \cite{Dubbers11}.

%\paragraph{Vud:}
The transition amplitude $V_{ud}$ of the CKM matrix \cite{Cabibbo63,Kobayashi73} can be derived using our result Eq.~\eqref{eq:final-result} and the updated world-average of the neutron lifetime measurements \cite{Dubbers18}, which includes the new measurements \cite{Serebrov18,Pattie18,Ezhov18}
\begin{equation}
\tau_n = \unit{879.7(8)}{s} \label{eq:lifetime}
\end{equation}
using Eq.~(2) from \cite{Czarnecki18}:
\begin{align}
 V_{ud}&=\left(\frac{4908.6(1.9)}{\tau_\mathrm{n}\cdot \left(1+3\lambda^2\right)}\right)^{1/2}\nonumber\\
 &=0.973 51 (19)_\mathrm{RC} (44)_{\tau_n} (35)_\lambda, \nonumber\\
 &=0.973 51 (60),
 \label{eq:vud}
\end{align}
where the error denoted RC stems from radiative corrections \cite{Marciano06}. This neutron result is in agreement with the average result from super-allowed beta decays of $V_{ud} = 0.974 17 (21)$ \cite{Hardy15} and only $2.9$ times less precise, see also \cite{Dubbers18}. We note that here we do not use the new common radiative correction with reduced uncertainty from \cite{Seng18} as theoretical discussions are ongoing.

% Tensor interactions
As a first application of our result to searches for new physics, we derive an improved limit on hypothetical left-handed tensor interactions.  Following the scheme of \cite{Pattie13,Pattie15} and using our result of the beta asymmetry Eq.~\eqref{eq:final-result}, our average of the neutron lifetime Eq.~\eqref{eq:lifetime} and the average nuclear $\mathcal{F}t$ value \citep{Hardy15}, we obtain:
\begin{equation}
-0.0044 < C_\mathrm{T} / C_\mathrm{A} < 0.00023, \quad (95\%\ \mathrm{C.L.}).
\end{equation}

\begin{acknowledgments}
The authors acknowledge the excellent support of the services of the Physikalisches Institut, Heidelberg University, and the ILL.  We thank D.~Rechten, TUM, for providing us with ultra-thin carbon foils, and U.~Schmidt, Heidelberg, for countless valuable discussions.
% Funding
This work was supported by the Priority Programme SPP~1491 of the German Research Foundation (DFG), contract nos. MA~4944/1-2, AB~128/5-2, SO 1058/1-1, the Austrian Science Fund (FWF) contract no. P~26636-N20, the German Federal Ministry for Research and Education, contract nos. 06HD153I and 06HD187, and the DFG cluster of excellence `Origin and Structure of the Universe'. The computational results presented have been achieved in part using the Vienna Scientific Cluster (VSC).
\end{acknowledgments}


%\bibliography{theory,review,correlations,polarisation,lifetime,Perkeo3A,nuclear}

\end{document}